\documentstyle[twoside,fleqn,espcrc2,epsf]{article}

\def\bbox#1{\mbox{\protect\boldmath $#1$}}
\def\case#1#2{{\textstyle \frac{#1}{#2}}}
\def\text#1{{\rm #1}}

\def\lsim{\mathop{\raisebox{-.4ex}{\rlap{$\sim$}} \raisebox{.4ex}{$<$}}}

\def\vek#1{\mbox{\protect\boldmath $#1$}}

\newcommand{\AmS}{{\protect\the\textfont2
  A\kern-.1667em\lower.5ex\hbox{M}\kern-.125emS}}

\title{Perturbative and Non-perturbative Corrections to
$B\to D^{(*)}l\nu$%
\thanks{Presented at Lattice '99, 28 June--3 July 1999, Pisa, Italy}
}

\author{Andreas S. Kronfeld%
	\address{Theoretical Physics Group, 
	Fermi National Accelerator Laboratory, 
	Batavia, Illinois, U.S.A.}%
\hfill\normalsize FERMILAB-CONF-99/251-T%
}

\begin{document}

\begin{abstract}
It is shown that certain double ratios introduced for computing 
semileptonic form factors are accurate to order $1/m_Q^2$, even when 
the action and current are accurate to order~$1/m_Q$.
\end{abstract}

\maketitle

\section{INTRODUCTION}
\label{sect:intro}

The semileptonic decays $B\to Dl\nu$ and $B\to D^*l\nu$ are crucial 
to the determination of the entry~$|V_{cb}|$ in the CKM matrix.
Experiments measuring the differential rate of these processes require 
theoretical input to extract~$|V_{cb}|$, namely the form factors of 
the hadronic transition.

To calculate the form factors we have introduced several double 
ratios~\cite{Hashimoto:1999yp,Sim99}
\begin{equation}
	R_+ = \frac{\langle D|V_0|B\rangle \langle B|V_0|D\rangle}{
		\langle D|V_0|D\rangle \langle B|V_0|B\rangle} =
		\rho_{V^{cb}_0}^{-2}|h_+|^2;
	\label{eq:R+}
\end{equation}
a similar ratio $R_1$ defined by replacing the pseudoscalars~$B$ 
and~$D$ with vectors~$B^*$ and~$D^*$;
\begin{equation}
	R_- = \frac{\langle D|V_i|B\rangle}{\langle D|V_0|B\rangle}
		\frac{\langle D|V_0|D\rangle}{\langle D|V_i|D\rangle} = 
		\rho_{V^{cb}_i}^{-1}\left[1-\frac{h_-}{h_+}\right];
	\label{eq:R-}
\end{equation}
and
\begin{eqnarray}
	R_A &\!\!=\!\!& 
		\frac{\langle D^*|\epsilon\cdot A|B\rangle 
		\langle B^*|\epsilon\cdot A|D\rangle}{
		\langle D^*|\epsilon\cdot A|D\rangle
		\langle B^*|\epsilon\cdot A|B\rangle} \nonumber \\
		&\!\!=\!\!&
		\rho_{A^{cb}}^{-2}
		\frac{h_{A_1}^{B\to D^*}h_{A_1}^{D\to B^*}}{
		      h_{A_1}^{D\to D^*}h_{A_1}^{B\to B^*}}.
	\label{eq:RA}
\end{eqnarray}
The ratios $R_+$, $R_1$, and~$R_A$ are defined at zero recoil;
the ratio~$R_-$ is defined in the limit of zero recoil.
The~$\rho$ are matching factors, needed to patch radiative 
corrections from short distances.

The ratios $R_+$ and $R_-$ directly give the form factors~$h_+$ 
and~$h_-$, which together form the hadronic amplitude for $B\to Dl\nu$.
Information from~$R_+$, $R_1$, and~$R_A$ must be extracted from their 
heavy quark expansions to obtain $h_{A_1}^{B\to D^*}$, the hadronic 
amplitude for $B\to D^*l\nu$~\cite{Sim99}.

In the limit of degenerate heavy quarks and in the infinite mass limit,
all four double ratios are equal to one.
Thus, one essentially computes the deviation of the ratios from one, 
and the statistical and systematic uncertainties are a fraction of 
$R-1$, not of~$R$.
This makes it possible to extract the $1/m_Q$ correction to~$h_-$ and
the $1/m_Q^2$ corrections to $h_-$, $h_+$, $h_1$, and $h_{A_1}$,
provided the action and currents are accurate enough.
The aim of this paper is to explain a remarkable result:
the double ratios yield the $1/m_Q^2$ corrections when the action and 
currents used to compute them are tuned only through order~$1/m_Q$.

\section{POWER CORRECTIONS}
\label{sect:power}

Properties of heavy quark states calculated with Wilson fermions can 
be interpreted by appealing to a non-relativistic effective theory, 
which provides a ``factorization'' of short-distance from 
long-distance physics~\cite{KKM97}.
As in any effective theory, the effects of short distances (here $a$, 
$m_b^{-1}$, and $m_c^{-1}$) are lumped into coefficients, while the 
effects of long distances (here $\Lambda_{\rm QCD}^{-1}$) are 
generated by local operators.

To deduce the operators of this effective theory, one can start 
by thinking about symmetries.
The action for Wilson fermions can be written
\begin{equation}
	S = \sum_x\bar{\psi}_x\psi_x 
	  - \kappa \sum_{x,y}\bar{\psi}_x M_{xy}\psi_y,
	\label{eq:S kappa}
\end{equation}
where $\kappa$ is the hopping parameter and the hopping 
matrix~$M_{xy}$ may include a clover term and further improvement 
terms.
The heavy-quark limit corresponds to small~$\kappa$, and as~$\kappa\to 
0$ the lattice action~(\ref{eq:S kappa}) obviously acquires the spin 
and flavor symmetries~\cite {Isgur:1989vq} of continuum QCD in the 
limit $\Lambda_{\rm QCD}/m_Q\to 0$.
As long as $\Lambda_{\rm QCD}/m_Q\ll 1$, Green functions calculated 
from~(\ref{eq:S kappa}) are given by the static limit plus small 
corrections, no matter what value~$m_Qa$ takes.

Thus, just as continuum QCD can be described by a heavy-quark 
effective theory (HQET), lattice QCD can be described by a modified 
HQET, as long as the relevant physical momenta~$\vek{p}$ satisfy
\begin{equation}
	|\vek{p}|\ll m_Q, \quad |\vek{p}|\ll 1/a.
	\label{eq:small p}
\end{equation}
The operators of the HQET for lattice QCD are the same as always, but 
the coefficients are different: they depend on ratios of short 
distances: $am_b$, $am_c$, and $m_b/m_c$.
The operators are sensitive to the scale~$\Lambda_{\rm QCD}$.
Heavy-light matrix elements may have lattice artifacts, but they arise 
from the light sector and are order~$(a\Lambda_{\rm QCD})^n$.
Lattice artifacts from the heavy quarks are absorbed into the 
coefficients, which deviate from their continuum limits; they can be 
quantified by computing the coefficients, say in perturbation theory 
in~$g_0^2$, and using tools of the HQET to propagate them to matrix 
elements.

Through order~$1/m_Q^2$ the action of the HQET for lattice QCD can be 
written
\begin{equation}
	S = \int d^{4}x\,\left[
		\bar{h}(D_t + m_1)h 
	  -	{\cal L}^{(1)} - {\cal L}^{(2)}\right],
	\label{eq:Seff}
\end{equation}
where $h=\gamma_0h$ is the heavy-quark field of the usual HQET, and 
all interactions ${\cal L}^{(n)}$ have the Dirac structure~$1$, 
$\gamma_0$, or~$\vek{\Sigma}$.
The higher-dimension interactions are
\begin{eqnarray}
	{\cal L}^{(1)} & = & 
		\frac{\bar{h}\vek{D}^{2} h}{2m_2} + 
		\frac{i\bar{h}\vek{\Sigma}\cdot\vek{B} h}{2m_B} 
	\label{eq:L1}  \\
	{\cal L}^{(2)} & = & \frac{\bar{h}[
		\vek{\gamma}\cdot\vek{D},
		\vek{\gamma}\cdot\vek{E}] h}{8m_E^2} +O(g_0^2),
	\label{eq:L2}
\end{eqnarray}
and there are more at even higher dimensions.
For example, at dimension~7 one finds
${\cal L}^{(3)} = \cdots + w_4\bar{h}\sum_{i=1}^3 D_i^4 h$.
Rotational invariance of continuum QCD implies $w_4=0$ in the usual 
HQET.
In the HQET describing lattice QCD, however, $w_4$ does not vanish 
unless the lattice action has been improved accordingly.

To describe matrix elements of (the lattice theory's) currents in the
(the lattice theory's) HQET, one must introduce effective currents.
For the vector current, for example,
\begin{equation}
	Z_{V^{cb}}V^{cb}_\mu \mapsto \eta_V \bar{h}'\gamma_\mu h 
		+ V_\mu^{(1)} + V_\mu^{(1,1)} + V_\mu^{(2)},
	\label{eq:Jeff}
\end{equation}
where
\begin{eqnarray}
V^{(1)}_\mu & = & 
	\frac{(\vek{D}\bar{h}')\cdot\vek{\gamma}\gamma_\mu h}{2m_{3c}} - 
	\frac{\bar{h}' \gamma_\mu \vek{\gamma}\cdot\vek{D} h}{2m_{3b}} 
		\label{eq:J1}  \\
V^{(1,1)}_\mu & = & - C_V^{(1,1)}
	\frac{(\vek{D}\bar{h}')\cdot\vek{\gamma}\gamma_\mu
	 	\vek{\gamma}\cdot\vek{D} h}{4m_{3c}m_{3b}} 
		\label{eq:J11}  \\
V^{(2)}_\mu & = & 
	\frac{(\vek{D}^{2}\bar{h}') \gamma_\mu h}{8m_{D_\perp^2c}^2} + 
	\frac{i\bar{h}'\vek{\Sigma}\cdot\vek{B} \gamma_\mu h}{8m_{\sigma Bc}^2}
		\nonumber \\ & + &
	\frac{ \bar{h}'\vek{\alpha}\cdot\vek{E} \gamma_\mu h}{4m_{\alpha Ec}^2} 
		\label{eq:J2}
\end{eqnarray}
In $V^{(1,1)}_\mu$, the coefficient $C_V^{(1,1)}=1+O(\alpha)$. 
In $V^{(2)}_\mu$, the $1/m_b^2$ terms are not written out, 
but they should be clear from the $1/m_c^2$ terms.

The basis of operators used in (\ref{eq:L1})--(\ref{eq:J2}) 
% (\ref{eq:L2}), (\ref{eq:J1}), and (\ref{eq:J2}) 
is not used in all papers on the usual HQET.
Because it avoids operators that ``vanish by the equations of motion,''
it is convenient for computing the radiative corrections with the
method of sect.~\ref{sect:loops}.
Other bases in the literature are related to this one by field 
redefinitions.

The rest mass~$m_1$ and the inverse ``masses'' $1/m_2$, $1/m_B$, 
$1/m_3$, $1/m_E^2$, etc., are the modified coefficients.
They depend both on couplings of the lattice action, notably on the 
bare quark mass and the gauge coupling.
The first two coefficients $m_1$ and $1/m_2$ are known to one 
loop~\cite{Mer98}.

In the asymptotic continuum limit, $m_Qa\to 0$, all coefficients 
obtain the same value as in the usual HQET.
In practice, however, everyone's Monte Carlo calculation falls roughly 
in the range
\begin{equation}
	\case{1}{2} \lsim m_ba \lsim 2,
	\label{eq:mb range}
\end{equation}
which is not, in any sense, asymptotic.
The point of the modified HQET is that in the range~(\ref{eq:mb range}) 
it is ideally suited for propagating the heavy quarks' discretization 
effects to hadronic matrix elements.

The rest mass~$m_1$ does not propagate to observables for a simple 
reason.
In the Hamiltonian formalism of the effective theory, the rest mass 
operator $m_1\bar{h}h$ commutes with all other terms in the 
Hamiltonian.
Eigenstates of the HQET are independent of~$m_1$, and mass dependence 
of the full eigenstates is acquired only from~${\cal L}^{(n)}$.

The power corrections to the symmetry limit the matrix elements 
in~(\ref{eq:R+})--(\ref{eq:RA}) are computed by treating the 
${\cal L}^{(n)}$ as perturbations.
Through order~$1/m_Q^2$ one must consider
$\bar{h}'\gamma_\mu h$,
$V^{(1)}_\mu$,
$T\{V^{(1)}_\mu {\cal L}^{(1)}\}$, 
$V^{(1,1)}_\mu$,
$T\{{\cal L}^{(1)} \bar{h}'\gamma_\mu h {\cal L}^{(1)}\}$,
$V^{(2)}_\mu$, and $T\{\bar{h}'\gamma_\mu h {\cal L}^{(2)}\}$,
where $T$ is the time-ordering symbol.
The correction $T\{\bar{h}'\gamma_\mu h {\cal L}^{(1)}\}$ vanishes 
by Luke's theorem~\cite{Luke:1990eg}.
One must now repeat, in the present basis of operators, calculations 
in Refs.~\cite{Falk:1993wt,Mannel:1994kv}, keeping track of all the 
inverse masses~\cite{Kronfeld:1995nu}.
The full results will be presented elsewhere.

Here we give a simple argument why the terms $V^{(2)}_\mu$ and 
$T\{\bar{h}'\gamma_\mu h {\cal L}^{(2)}\}$ drop out of the double 
ratios.
In the HQET's normalization, currents are 1 plus corrections.
To order~$1/m_Q^2$, the offending terms factor.
Taking $T\{\bar{h}'\gamma_\mu h {\cal L}^{(2)}\}$ first 
\begin{equation}
	\langle D|J|B\rangle = (1 + \Theta_c/m_{Ec}^2)(1 + \Theta_b/m_{Eb}^2),
	\label{eq:DJB}
\end{equation}
where $\Theta_c$ and $\Theta_b$ are unknowns.
As expected, $T\{\bar{h}'\gamma_\mu h {\cal L}^{(2)}\}$, and similarly 
$V^{(2)}_\mu$, {\em does} effect the 
individual matrix elements, but after inserting~(\ref{eq:DJB}) 
into~(\ref{eq:R+}) or~(\ref{eq:RA}) they cancel.

The only effects of order~$1/m_Q^2$ which survive are of the form
\begin{equation}
	\langle D|J|B\rangle = 1 + \lambda/(m_cm_b),
	\label{eq:DVB}
\end{equation}
for matrix elements, becoming 
\begin{equation}
	R=1-\lambda(1/m_c-1/m_b)^2
\end{equation}
in~(\ref{eq:R+}) and~(\ref{eq:RA}).
In~$R_+$, $R_1$, and $R_A$ these effects arise from $V_{\mu}^{(1,1)}$
and $T\{{\cal L}^{(1)} \bar{h}'\gamma_\mu h {\cal L}^{(1)}\}$.
Neglecting the radiative correction $C_J^{(1,1)}-1$, these double
ratios---and so~$h_+$, $h_1$, and $h_{A_1}$---have the right mass
dependence if $m_2=m_B=m_3$.

The terms $V_\mu^{(1)}$ and $T\{V_\mu^{(1)} {\cal L}^{(1)}\}$ make 
contributions only to $\langle D|V_i|B\rangle$, so~(\ref{eq:R-}) 
gives~$h_-$ the right mass dependence also, if $m_3=m_2=m_B$.

\section{RADIATIVE CORRECTIONS}
\label{sect:loops}

To compute the radiative corrections in perturbation theory, one 
calculates matrix elements of quark states.
Taking into account Gordon identities (of the lattice spinors)
\begin{equation}
	\hspace*{-0.1cm}
	\begin{array}{l}
	Z_{V^{cb}}\langle c,\bbox{p}'|V^\mu|b,\bbox{p}\rangle =
		\bar{u}'\gamma^\mu u \, F_\mu + \\
	\hspace*{0.1cm}
   		\bar{u}'i\sigma^{\mu\nu}u [
   			i(v'-v)_\nu H_{\mu\nu}^{(+)} +
			i(v'+v)_\nu H_{\mu\nu}^{(-)}], 
	\end{array}
	\label{eq:cVb}
\end{equation}
(sum on $\nu$, but not on $\mu$).
The velocities satisfy ${\kern+0.1em /\kern-0.55em v}u=iu$ 
and~$v^2=-1$.
The functions~$F_\mu$ and~$H_{\mu\nu}^{(\pm)}$ are evaluated on 
(lattice) mass shell, and expanded in~$\bbox{p}$ and~$\bbox{p}'$.

The factor~$Z_{V^{cb}}$ is chosen to yield the radiative 
corrections of the continuum at~$\bbox{p}'=\bbox{p}=\bbox{0}$, viz.\
\begin{equation}
	\left. F_0\right|_{\rm on~shell}=\eta_V.
\label{eq:F0=1}
\end{equation}
Similarly, the rotation parameter~$d_1$ \cite{KKM97} should be 
adjusted so that $F_i=F_0$ on shell.

The factor~$Z_{V^{cb}}$ has strong mass dependence, 
$Z_{V^{cb}}\sim e^{(m_1^c+m_1^b)/2}$, and its (bare) perturbative 
series has large coefficients from tadpole diagrams.
In matching factors~$\rho$ needed for the double ratios both 
vices cancel: in~(\ref{eq:R+}) and~(\ref{eq:RA})
\begin{equation}
	\rho_{V^{cb}_0} = \frac{Z_{V^{cb}}Z_{V^{bc}}}{Z_{V^{cc}}Z_{V^{bb}}},
		\quad
	\rho_{A^{cb}}   = \frac{Z_{A^{cb}}Z_{A^{bc}}}{Z_{A^{cc}}Z_{A^{bb}}},
	\label{eq:rhoV0}
\end{equation}
where~$Z_{A^{cb}}$ is defined by a condition similar to~(\ref{eq:F0=1}),
and in~(\ref{eq:R-})
\begin{equation}
	\rho_{V^{cb}_i} =
		\frac{F_i(m_c,m_b) + 2H_{i0}^{(-)}(m_c,m_b)}{F_i(m_c,m_c)}.
	\label{eq:rhoVi}
\end{equation}
The factors~$\rho$ vary smoothly from the continuum limit (where they 
equal~1) to the static limit.
At one loop $\rho-1$ prove to be small~\cite{Kro98}.

\vspace{2.875mm}
Fermilab is operated by Universities Research Association Inc.,
under contract with the U.S. Deptartment of Energy.

\end{document}